\documentclass[twocolumn,showpacs,preprintnumbers,amsmath,amssymb]{revtex4}
\usepackage{graphicx}
\usepackage{dcolumn}
\usepackage{bm}
\usepackage{subfigure}

\begin{document}
\preprint{}
\title{Disease spreading in populations of moving agents}

\author{Arturo Buscarino$^1$}
\author{Luigi Fortuna$^1$}
\author{Mattia Frasca$^1$}
\author{Vito Latora$^2$}
\affiliation{$^1$Dipartimento di Ingegneria Elettrica, Elettronica
e dei Sistemi, Universit\`a degli Studi
di Catania, viale A. Doria 6, 95125 Catania, Italy\\
$^2$Dipartimento di Fisica e Astronomia, Universit\`a di Catania,
and INFN, Sezione di Catania, Via S. Sofia, 64, 95123 Catania,
Italy}

\date{\today}
\begin{abstract}
We study the effect of motion on disease spreading in a system 
of random walkers which additionally perform long-distance jumps. 
A small percentage of jumps in the agent motion is sufficient to destroy 
the local correlations and to produce a large drop in the epidemic 
threshold, well explained in terms of a mean-field approximation. 
This effect is similar to the crossover found in static small-world 
networks, and can be furthermore linked to the structural 
properties of the dynamical network of agent interactions. 
\end{abstract}
\maketitle

Many information/communication and social systems 
can be modeled as complex networks 
\cite{barabasireview,newmanreview,boccalettireview}. 
One of the main reasons for studying such networks is to 
understand the mechanisms by which information, rumors and 
diseases spread over them. 
Recent works have pointed out the importance of
incorporating the peculiar topology of the underlying network in the
theoretical description of {\em disease spreading}
\cite{anderson92,murray,hethcote}.
Epidemic models are in fact heavily affected by the connectivity
patterns characterizing the population in which the infective
agent spreads. Both the nature of the final state, and the dynamics of
the disease process, strongly depend on the coupling topology.
Specifically, spreading occurs faster in {\em small-world} systems, i.e.
in networks with shorter characteristic path
lenghts \cite{wattsbook,kuperman}.
Moreover, the epidemic threshold is affected by the properties of
the degree distribution $P(k)$. For instance, the divergence of
the second-order moment of $P(k)$ leads, in
uncorrelated {\em scale-free} networks, to the surprising result
of the absence of an epidemic threshold and its associated critical
behavior \cite{pv00,moreno02,newman02}. This implies that scale-free
networks are prone to the spreading of infections
at whatever spreading rate the epidemic agents possess.

Most of the results present in the literature so far refer
to cases where the disease spreading takes place over a
wiring topology that is static, i.e. the underlying network is fixed
in time, or grown, once forever.
A more realistic possibility is to consider the networks themselves
as dynamical entities. This means that the topology is allowed
to evolve and adapt in time, driven by some external factors or by
the very same spreading process.
For instance, Refs.\cite{gross,bagnoli} have considered 
disease spreading on adaptive networks in which the susceptible 
agents have perception of the risk of infection, and 
are able to avoid contact with infected agents by rewiring 
their network connections. 
In this Letter we study disease spreading on a system of mobile
agents. The agents are random walkers which can additionally
perform long-distance jumps, and are only able to interact with
agents falling within a given interaction radius apart from them.
Hence, the interaction network between individuals is a dynamical
one, because the links evolve in time according to the agent movement. 
The focus of our work is on the influence of the motion on the disease
spreading. With the aim of large scale simulations, we will show
that the motion, usually neglected in epidemic models, has instead
a profound effect on the dynamics of the spreading and on the epidemic 
threshold. In particular, a small number of long-distance jumps in the agent 
motion is sufficient to produce a large drop in the epidemic threshold,  
as that observed in static small-world networks \cite{kuperman}. 
The case of infecting moving individuals has been considered only
in few other works \cite{Boccara92,gonzalez04,eubank04,colizza06,
frasca06,nekovee07}, 
although the effects on the epidemic threshold have never been studied 
in detail before. 
This is a quite important issue both in social and in artificial 
networks. E.g., recently the analogy with epidemic spreading has been
exploited to propose routing algorithms in highly mobile networks of
computers \cite{Neglia05,mascolo06}

We consider a system of $N$ identical agents independently
moving in a two-dimensional cell of linear size $D$,
with periodic boundary conditions. Fixing the value of $D$ is
equivalent to fix the agent density $\rho= N/D^2$.
The agents are represented as point particles, and their positions
and velocities at time $t$ are indicated as $\mathbf{r}_i(t)$ and
$\mathbf{v}_i(t) \equiv ( v_i(t) \cos\theta_i(t), v_i(t)
\sin\theta_i(t))$, $i=1,...,N$. We further impose that the agents
move with a velocity modulus which is constant in time
and equal for all the agents, i.e. $v_i(t)=v, \forall i=1,..,N$ 
and $\forall t$.
At time $t=0$ the $N$ particles were distributed at random.
At each time step, the agents change stochastically the direction
angles $\theta_i(t)$. The positions and the orientations of the particles 
are thus updated according to the following rule:
\begin{eqnarray}
  \theta_i(t) &=&\xi_i
\nonumber
\\
 \mathbf{r}_i(t+1)     &=& \mathbf{r}_i(t)+\mathbf{v}_i(t)
\label{eq:move}
\end{eqnarray}
where $\xi_i$ are $N$ independent identically distributed random
variables chosen at each time with uniform probability in the
interval $[-\pi,\pi]$. In addition, to include the possibility
that agents can move through the bidimensional world with time
scales much shorter than those related to disease, as in the
case of infected individuals travelling by flights \cite{colizza06}, 
we consider that agents can perform long-distance jumps. This is
accounted for by defining a parameter, $p_{j} \in [0,1]$, that
quantifies the probability for an agent to perform a jump into a
completely random position. In summary, at each time step, each
agent evolves: following Eqs.~(\ref{eq:move}), with a probability $1-
p_{j}$, or performing a jump, with probability $p_{j}$. In the latter 
case the position of the agent is updated into a new position
chosen at random in the cell. 
Models with different jumping rules \cite{frasca06},
non-indentical and interacting agents, have also been considered,
and the results will be reported elsewhere. Finally, the main
parameters controlling the moving agents in our model are: $\rho$,
$v$ and $p_{j}$.

Among the possible mechanisms of disease spreadings
\cite{anderson92,murray,hethcote},
we focus on the {\em SIR} model, that divides the $N$ agents into three
disjoint groups: susceptible (S), infective (I) and recovered (R).
We indicate as $N_S(t)$, $N_I(t)$ and $N_R(t)$, respectively, the number
of agents in the three  groups at time $t$, with the total number of agents 
$N_S(t)+N_I(t)+N_R(t) = N$ being constant in time. 
A small number of agents is set in the infective state 
at $t=0$ as the seed of the infection,
while all the others start from the susceptible state.
The process through which the disease spreads can be
summarized as following. An interaction radius $r$ 
is fixed ($r=1$ in all our calculations), and this defines the
interaction network: at each time step $t$ each agent interacts 
only with those agents located within a neighborhood of radius $r$.
For a given susceptible agent, the probability of being 
infected increases with
the number of infected individuals in the neighborhood. More
precisely, if an agent is in the $S$ state at time $t$, and
exactly one of its neighbors is in the $I$ state, then it moves
into the $I$ state with probability $\lambda$
and stays in the $S$ state with probability $1-\lambda$.
If $N_{I_r}$ is the number of infected individuals in
the neighborhood of the agent, then its probability of being
infected is $1-(1-\lambda)^{N_{I_r}}$. In addition to this, 
each infected agent can move into the $R$ state with probability $\mu$, 
and then cannot catch the disease anymore. This sets the 
average duration time of the infection: $\tau=\frac{1}{\mu}$.

\begin{figure}[h]
\centering {{\includegraphics[width=9cm]{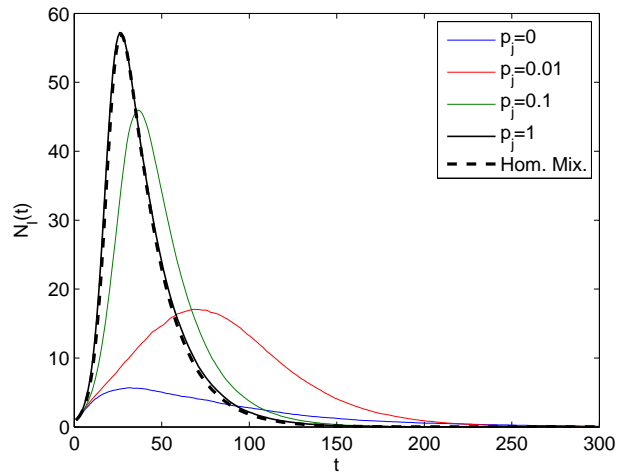}}}
\caption{ (Color online). Number of infected individuals as a
function of time $t$. We have considered a system with $N=1000$,
$\rho=1$, $v=0.1$, $\lambda=0.1$, $\mu=0.05$ and different values
of $p_{j}$. The system is started with $1\%$ of the agents set in
the infective states. Results are averages over 100 runs. The
dashed line is the result in the homogeneous mixing approximation of 
Eqs.~(\ref{eq:modelloHomMixing})} 
\label{fig:dynamics}
\end{figure}

In our model we implement the motion 
rules and we update, at each time step, the disease state of every agent. 
The model is simulated for a number of time steps
sufficiently high to ensure that, at the end, there are no
more infected individuals in the population.
During a simulation, the number $N_I(t)$ of infected grows up, reaches
a peak value, and then decreases. Typical cases are shown in
Fig.~\ref{fig:dynamics}, 
where it can be noticed that a larger value of $p_{j}$ increases
the spread of the infection. In fact, the peak
of infected individuals is higher in the presence
of a larger probability of jumping. We have verified that this is
also true for the total number of individuals which have
contracted the disease at the end of the process. Both the two
issues have important practical consequences, since on one hand
the disease involves a higher percentage of the population, and,
on the other hand, it requires more resources to deal with a
higher peak of infected individuals.

The behavior for large $p_j$ and/or large $v$ can be interpreted 
in terms of a mean-field approximation. 
In fact, in such limit, we expect that the spatial correlations in the
disease states are destroyed by the agent motion, and we assume
that the {\em homogeneous mixing (HM)} hypothesis is valid.
Under this hypothesis, all the individuals have the  
same probability of contacting any other individual 
(i.e., the population mixes at random) 
\cite{anderson92,murray,hethcote,Boccara92}, 
and the equations for the system of infectious mobile 
agents read:
\begin{eqnarray}
  i(t+1) &=& i(t)+ s(t) \left[1-\left(1-\lambda i(t)\right)^{a}\right]-\mu i(t)
\nonumber
\\
  r(t+1) &=& r(t)+ \lambda i(t)
\label{eq:modelloHomMixing}
\\
  s(t+1) &=&  \rho - i(t+1) - r(t+1)
\nonumber
\end{eqnarray}
where $s(t)=N_S(t)/D^2$, $i(t)=N_I(t)/D^2$, $r(t)=N_R(t)/D^2$, are
respectively the densities of susceptible,
infected and recovered individuals at time $t$,
and $a=\pi r^2$. The third equation is simply derived from the
conservation of the number of agents.
The second equation indicates that the increase of recovered individuals
at time $t+1$ is proportional to the number of infected individuals which
get recovered, i.e. to $\lambda i(t)$.
The first equation can be derived by taking into account that the
density of infected individuals at time $t+1$ is decreased by
$\lambda i(t)$ and increased by the density of susceptibles
catching the disease. This last term is proportional to $s(t)$
times a contagion probability $p_{cont}$. The contagion probability
is given by $p_{cont}=1-\bar{p}_{cont}$, where $\bar{p}_{cont}$
represents the probability of not being infected. $\bar{p}_{cont}$
is the probability that an agent will not being infected by none
of its neighbors, i.e. $\bar{p}_{cont}=\left(1-\lambda
\frac{N_I(t)}{D^2}\right)^{a}$, with $a$ representing the area in
which each agent may sense other infective individuals, that in our
case is equal to $\pi r^2$. The number of infective agents as a
function of the time in the HM approximation computed
from Eqs.~(\ref{eq:modelloHomMixing}), with 
$\lambda=0.1$ and $\mu=0.05$, is reported as a dashed line in 
Fig.~\ref{fig:dynamics}. 
As expected, the curves for the model approach the dashed
line when $p_j \rightarrow 1$.
The mean field approach gives us also information on the
epidemic threshold. In fact, for small $i(t)$ we can approximate
$\left(1-\lambda i(t)\right)^a \simeq 1-a \lambda i(t)$, and  we
get from Eqs.~(\ref{eq:modelloHomMixing}) the iterative rule
$i(t+1)= i(t)+ \pi r^2 \lambda s(t) i(t)-\mu i(t)$.
This allows to distinguish two cases. In fact, by assuming
$s(0) \simeq \rho$, we get $i(1)>i(0)$ when
$\sigma \equiv \lambda/\mu > \frac{1}{\pi r^2 \rho}$, while
$i(1)<i(0)$ when $\sigma \equiv \lambda/\mu < \frac{1}{\pi r^2 \rho}$.
Thus, under the HM hypothesis we derive a critical
threshold:
\begin{equation}
 \sigma_c = \frac{1}{\pi r^2 \rho}
 \label{eq:soglialcvsrho}
\end{equation}
Hence, when $\sigma < \sigma_c$, the number of infected decrease monotonically,
while for $\sigma > \sigma_c$ an epidemic outbreak occurs.
Also notice that $\sigma_c = \frac{1}{<k>}$ as found in
Erd\H{o}s and R\'enyi random graphs \cite{newmanreview,boccalettireview},
where $<k>$ is the average number of first neighbours of an agent,
that in our case is equal to $\rho\pi r^2$.
\begin{figure}[h]
\centering
{{\includegraphics[width=9cm]{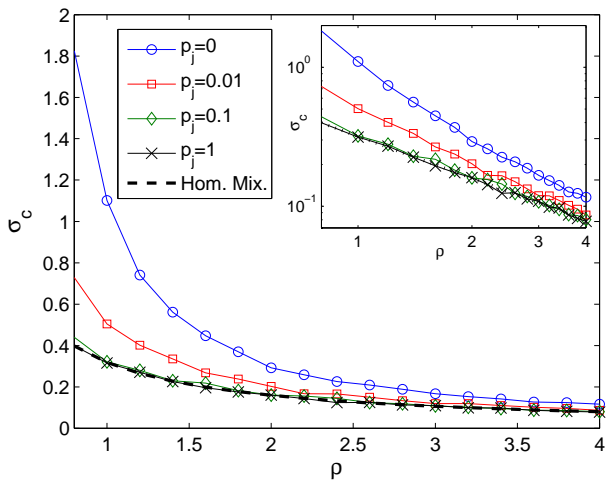}}}
\caption{(Color online). Epidemic threshold as a function of the
system density $\rho$. We have considered $N=1000$, $v=0.1$, and
$\mu=0.05$.
Different curves correspond to different values of $p_{j}$.
Results are averages over 100 runs.
The dashed line is the prediction in the homogeneous mixing approximation.}
\label{fig:sigmavsrho}
\end{figure}

In Fig.~\ref{fig:sigmavsrho} we report the epidemic thresholds $\sigma_c$ 
computed numerically for the model of infective agents moving with different 
values of $p_j$. For a given density $\rho$, 
We observe that  $\sigma_c$ is a decreasing function of the density 
$\rho$. Moreover, for a given value of $\rho$, the threshold decreases 
with the jumping probability $p_j$. 
In the same figure we report for comparison the prediction of 
Eq.~(\ref{eq:soglialcvsrho}), as a dashed line. 
We notice that the homogeneous mixing approximation becomes 
more and more accurate when $p_{j}$ tends to 1. The convergence to 
the HM threshold of Eq.~(\ref{eq:soglialcvsrho}) as a function of $p_j$ 
is rather fast. For instance, already at $p_j = 0.1$, the threshold in the
model is, for any value of $\rho$ reported, practically
indistinguishable from the HM one. In Fig.~\ref{fig:sigmavsrho} we have 
considered a fixed velocity $v=0.1$. We have also studied 
$\sigma_c$ as a function of $v$, at a fixed density
and for different values of $p_j$. We observed that 
for small values of $v$, the epidemic threshold tends to the 
prediction of the HM when $p_j$ tends to 1, while, for large enough 
values of $v$, the epidemic threshold is consistent 
with the homogenous mixing one, independently of $p_j$.

\begin{figure}[h]
\centering
{{\includegraphics[width=9cm]{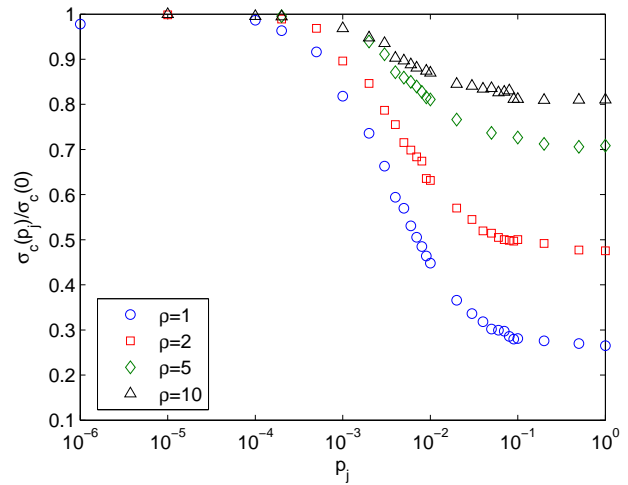}}}
\caption{\label{fig:sigmacvspj} (Color online). Scaled epidemic
threshold $\sigma_c (p_{j})/ \sigma_c (p_{j}=0)$ as a 
function of $p_{j}$. We have considered $v=0.1$ and $\mu=0.05$. 
Different curves correspond to different values of the density $\rho$. 
Results are averages over 100 runs.}
\end{figure}

Finally we investigate in more details the effects of $p_j$ on the 
epidemic threshold. In Fig.~\ref{fig:sigmacvspj} we report,  
as a function of $p_j$, the value of the threshold $\sigma_c(p_j)$,  
normalized by $\sigma_c(p_j=0)$, for different values of the density 
$\rho$. 
We observe a rapid drop in the curves 
(note the logarithmic scale for $p_j$), meaning that a small 
number of long-range jumps produce a large decrease in 
the epidemic threshold. The plateau observed for $p_j$ larger 
than $10^{-2}$ implies that in order to have a significant change 
in the epidemic threshold (and in the disease incidence) in our  
model of moving agents, the jumping probability has to be extremely 
small. This is in line with other results, stressing 
the role of the large scale properties of the airline 
transportation networks in determining the global diffusion 
pattern of emerging diseases \cite{colizza06}, and can have important 
implications  in the immunization of real communication networks 
\cite{gomez06}. 
Different curves in the figure correspond to different densities. 
As $\rho$ decreases, the drop in the curve occurs for 
smaller and smaller values of $p_j$, suggesting that no finite 
critical value of $p_j$ can be determined this way. 
This behavior is similar to the crossover observed in the
characteristic path length of small-world networks as a function 
of the rewiring probability \cite{wattsbook,Barrat00}. We notice, 
however, that in our case the effect is due to the agent movement 
and not to the rewiring of static links. 
\begin{figure}[h]
\centering {{\includegraphics[width=9cm]{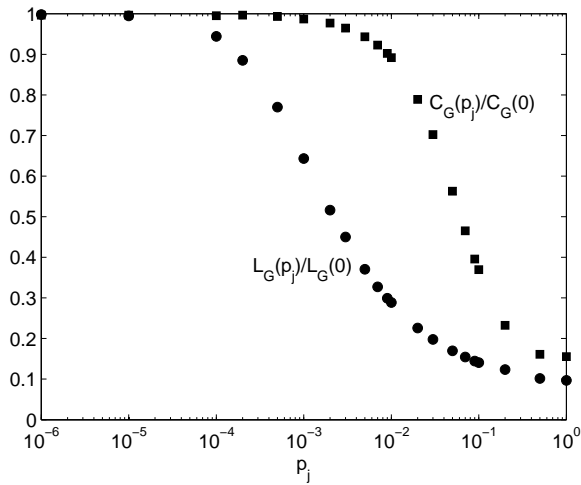}}}
\caption{\label{fig:LCvspj} Characteristic path
length $L_G$ (circles) and clustering coefficient $C_G$ (squares) 
as a function of $p_j$ ($N=1000$, $\rho=1$, $v=0.1$). 
Results are averages over 10 realizations.}
\end{figure}

The observed behaviour can be related to the topological 
properties of the underlying dynamical network. For such purpose  
we define an effective adjacency matrix $G_{\tau}(t)= \{ g_{ij}(t) \}$ 
taking into account that each infected individual may infect 
other individuals during the average duration of the
infection, i.e. during $\tau=\frac{1}{\mu}$ simulation 
steps \cite{frasca06}.  
Let $A(t)$ be the adjacency matrix at time $t$ defined so that
$a_{ij}(t)=1$ if the $j^{th}$ agent is within the interaction
radius of the $i^{th}$ agent at time $t$, and $a_{ij}(t)=0$
otherwise. We set $g_{ij}(t)=1$, if at least for
one $t'$, with  $t'= t, t-1, \cdots, t - \tau +1$, it is verified that
$a_{ij}(t')=1$. Otherwise we set $g_{ij}(t)=0$.
In Fig.~\ref{fig:LCvspj} we report the characteristic path length, $L_G$,   
and the clustering coefficient, $C_G$ of matrix $G(t)$ as a function 
of $p_j$. 
The behavior we observe is in all similar to that found in small-world 
networks for increasing rewiring probability \cite{wattsbook}. 
Notice that the drop in $L_G$ occurs at the same value of $p_j$ at 
which we have found the drop in the epidemic threshold.

In summary, we have considered a system of infectious mobile agents 
to study the effects of long-range moves on the disease spreading. 
Our results indicate that the interplay between dynamics and 
topology can have important consequences for the global spreading of 
infectious diseases in systems of mobile agents, and in the related 
applications such as the forecast of epidemic spreading  
and the development of wireless routing strategies.


\begin{thebibliography}{}



\bibitem{barabasireview} R. Albert and  A.-L.~Barab\'asi, Rev. Mod. Phys. {\bf 74}, 47 (2002).
\bibitem{newmanreview} M.E.J. Newman, SIAM Review {\bf 45}, 167 (2003).
\bibitem{boccalettireview} S. Boccaletti, V. Latora, Y. Moreno, M. Chavez and D -U Hwang, Phys. Rep.
{\bf 424} 175 (2006).
\bibitem{anderson92}  R.~M. Anderson and R.~M. May, {\em Infectious diseases in humans} (Oxford University Press, Oxford, 1992).
\bibitem{murray} J.~D. Murray, {\em Mathematical Biology} (Springer Verlag, Berlin, 1993).
\bibitem{hethcote} H.W. Hethcote, SIAM Review {\bf 42}, 599 (2000). 
\bibitem{wattsbook} D.J. Watts, {\it Small Worlds: The Dynamics of Networks between Order and Randomness},(Princeton University Press, Princeton, New Jersey, 1999).
\bibitem{kuperman} M. Kuperman and G. Abramson, Phys. Rev. Lett. {\bf 86}, 2909 (2001).
\bibitem{pv00} R.~Pastor-Satorras and A.~Vespignani. Phys. Rev. Lett. {\bf 86}, 3200 (2001).
\bibitem{moreno02}  Y.~Moreno, R.~Pastor-Satorras, and A.~Vespignani, Eur. Phys. J. {\bf B26}, 521 (2002). 
\bibitem{newman02} M.~E.~J. Newman, Phys. Rev. {\bf E66}, 016128 (2002). 
\bibitem{gross} T. Gross, C. J. D. D'Lima and B. Blasius,  Phys. Rev. Lett. {\bf 96}, 208701 (2006)
\bibitem{bagnoli} F. Bagnoli, P. Li\'o and L. Sguanci,  arXiv:0705.1974; L. Sguanci, P. Li\'o and F. Bagnoli,  arXiv:q-bio/0607010. 
\bibitem{Boccara92} N. Boccara and K. Cheong, J. Phys. {\bf A25}, 2447 (1992) 
\bibitem{gonzalez04} M.C. Gonzalez and H. J. Herrmann, Physica {\bf A340} 741 (2004) 
\bibitem{eubank04} S. Eubank, H. Guclu, V. S. A. Kumar, M. V. Marathe, A. Srinivasan, Z. Toroczkai and N. Wang, Nature {\bf 429}, 180 (2004)
\bibitem{colizza06} V. Colizza, A. Barrat, M. Barthelemy and A. Vespignani, 
Proc. Natl. Acad. Sci. {\bf 103}, 2015 (2006). 
\bibitem{frasca06} M. Frasca, A. Buscarino, A. Rizzo, L. Fortuna and S. Boccaletti, Phys. Rev. {\bf E74}, 036110 (2001).
\bibitem{nekovee07} M. Nekovee, New Journal of Physics {\bf 9}, 189 (2007).
\bibitem{Neglia05} E. Zhang, G. Neglia, J. Kurose and D. Towsle, Computer Networks {\bf 51}, 2867 (2007). 
\bibitem{mascolo06} M. Musolesi and C. Mascolo, in Proceedings of Int. Conference on Mobile and Ubiquitous Systems: Networks and Services (MOBIQUITOUS 2006). 
San Jose, CA. July 2006. ACM. 
\bibitem{gomez06} J. Gomez-Gardenes, P. Echenique and Y. Moreno,  Eur. Phys. J. {\bf B49}, 259 (2006).
\bibitem{Barrat00} A. Barrat, M. Weigt, Eur. Phys. J. {\bf B13}, 547 (2000).






\end{thebibliography}
\end{document}